\documentclass[aps,reprint,superscriptaddress,amsmath,amssymb]{revtex4-2}

\usepackage{graphicx}
\graphicspath{{figures/}}
\usepackage{siunitx}
\usepackage{braket}
\usepackage{xcolor}
\usepackage{booktabs}
\usepackage{array}
\usepackage{tabularx}
\usepackage{multirow}
\usepackage[bookmarks=false]{hyperref}
\usepackage[capitalise]{cleveref}
\crefname{section}{Sec.}{Secs.}
\Crefname{section}{Section}{Sections}

\definecolor{pink}{RGB}{255,0,255}

\definecolor{blue}{rgb}{0,0,1}

\definecolor{red}{rgb}{1,0,0}

\newcolumntype{P}[1]{>{\centering\arraybackslash}p{#1}}
\newcolumntype{L}{X}
\newcolumntype{C}{>{\centering\arraybackslash}X}
\newcolumntype{R}{>{\raggedright\arraybackslash}X}

\begin{document}

\title{Reversible photorefractive attack in continuous-variable quantum key distribution}

\affiliation{College of Electronic Science and Technology, National University of Defense Technology, Changsha 410073, China}
\affiliation{State Key Laboratory of Photonics and Communications, Institute for Quantum Sensing and Information Processing, Shanghai Jiao Tong University, Shanghai 200240, China}
\affiliation{College of Computer Science and Technology, National University of Defense Technology, Changsha 410073, China}
\affiliation{Shanghai Research Center for Quantum Sciences, Shanghai 201315, China}
\affiliation{Hefei National Laboratory, Hefei 230088, China}

\author{Junxuan Liu}
\affiliation{College of Computer Science and Technology, National University of Defense Technology, Changsha 410073, China}

\author{Yuchao Liu}
\affiliation{State Key Laboratory of Photonics and Communications, Institute for Quantum Sensing and Information Processing, Shanghai Jiao Tong University, Shanghai 200240, China}

\author{Qingquan Peng}
\affiliation{College of Electronic Science and Technology, National University of Defense Technology, Changsha 410073, China}

\author{Piao Tan}
\affiliation{State Key Laboratory of Photonics and Communications, Institute for Quantum Sensing and Information Processing, Shanghai Jiao Tong University, Shanghai 200240, China}

\author{Tao Wang}
\affiliation{State Key Laboratory of Photonics and Communications, Institute for Quantum Sensing and Information Processing, Shanghai Jiao Tong University, Shanghai 200240, China}
\affiliation{Shanghai Research Center for Quantum Sciences, Shanghai 201315, China}
\affiliation{Hefei National Laboratory, Hefei 230088, China}

\author{Peng Huang}
\email{huang.peng@sjtu.edu.cn}
\affiliation{State Key Laboratory of Photonics and Communications, Institute for Quantum Sensing and Information Processing, Shanghai Jiao Tong University, Shanghai 200240, China}
\affiliation{Shanghai Research Center for Quantum Sciences, Shanghai 201315, China}
\affiliation{Hefei National Laboratory, Hefei 230088, China}

\author{Anqi Huang}
\email{angelhuang.hn@gmail.com}
\affiliation{College of Electronic Science and Technology, National University of Defense Technology, Changsha 410073, China}

\date{\today}

\begin{abstract}
Quantum key distribution (QKD) offers information-theoretic security, but device imperfections introduce implementation vulnerabilities. Experimental evidence of the security consequences of transmitter vulnerabilities in continuous-variable QKD (CV-QKD) remains limited. Here, we experimentally investigate a reversible photorefractive attack in a local-local-oscillator CV-QKD system. Moderate blue-light illumination reduces the modulation variance of a lithium-niobate IQ modulator to a stable value without appreciably increasing estimated excess noise. The variance recovers after illumination is removed. We elucidate the underlying physical mechanisms and develop an attack that exploits recovery to compromise modulation calibration. Receiver measurements confirm that the resulting calibration mismatch leads to overestimated channel transmittance and underestimated excess noise. Key rate calculations based on measured parameters yield a positive estimated rate when the correctly calibrated rate is nonpositive. Recovery can also make the preceding manipulation harder to identify through later inspection. This work provides the first system-level experimental evidence that a reversible photorefractive response can be exploited to compromise modulation calibration in CV-QKD. These findings highlight reversible changes in quantum transmitters as a practical security concern, motivating security analyses and protection strategies that account for both optical perturbations and subsequent recovery.
\end{abstract}

\maketitle


\section{Introduction}
\label{Sec:Introduction}

Quantum key distribution (QKD), based on the principles of quantum mechanics, enables the information-theoretically secure distribution of secret keys between communicating parties~\cite{RevModPhys.74.145,RevModPhys.81.1301,RevModPhys.92.025002}. Continuous-variable QKD (CV-QKD) encodes information in the quadratures of optical fields and retrieves it through coherent detection. In recent years, CV-QKD has demonstrated transmission over more than 200 km of optical fiber~\cite{PhysRevLett.125.010502} and 9.6 km of free space~\cite{hvhx-pdrn}. Benefiting from high-speed electro-optic modulation~\cite{Wang2022,Pi:23,Wang:25}, low-cost and high-efficiency coherent detection~\cite{RevModPhys.92.025002,10.1145/3524455}, and compatibility with existing optical communication infrastructures~\cite{Grosshans2003,RevModPhys.84.621,Wang_2024}, CV-QKD has become a promising approach for metropolitan-area quantum-secured communication networks. As transmission performance and system implementations continue to advance, the security of practical devices remains essential to its deployment.

Practical QKD devices may deviate from the idealized models used in security analyses, creating implementation vulnerabilities~\cite{PhysRevLett.107.110501,PhysRevLett.112.070503,Jain_2021,Hajomer2022,Liu:24,nwag174,Su2026,33cb-kvyp}. In transmitted-local-oscillator CV-QKD, the local oscillator (LO) travels through the quantum channel together with the signal, providing an adversary with opportunities to manipulate the measurement reference~\cite{PhysRevA.88.022339,PhysRevA.87.062313,PhysRevApplied.20.024073}. Local-local-oscillator (LLO) architectures generate the LO at the receiver and remove the attack surface associated with its transmission~\cite{PhysRevX.5.041009,PhysRevX.5.041010,Huang:15}. Measurement-device-independent CV-QKD addresses the trust placed in receiver measurement devices~\cite{PhysRevA.89.052301,Pirandola2015,Tian:22}. However, the transmitter prepares and encodes the quantum states, so imperfections in its components or their responses to external manipulation can alter the transmitted states and affect system security. Identifying transmitter vulnerabilities and assessing their security implications therefore remain essential to strengthening the practical security of CV-QKD.

Previous studies of CV-QKD transmitter security have examined performance changes in variable optical attenuators (VOAs) under laser-damage or photorefractive attack~\cite{PhysRevA.100.012313,Zhou:25,Mao2026}, as well as intrinsic imperfections of light sources and the laser-seeding attack~\cite{Huang2013,Zheng:19}. These studies mainly concern light generation and attenuation. Modulators warrant particular attention because they directly encode the quantum states used for key distribution. For these devices, studies of the Trojan-horse attack and modulation leakage have identified potential information leakage~\cite{Jain_2014,Jain_2021}, while component experiments and numerical analyses have explored security risks arising from photorefraction~\cite{PhysRevApplied.19.054052,PhysRevApplied.20.044013,Wang2025IPA}. However, existing investigations have largely relied on theoretical analysis, numerical simulation, or component characterization~\cite{PhysRevA.100.012313,Zhou:25,Mao2026,Huang2013,Zheng:19,Jain_2021,PhysRevApplied.19.054052,PhysRevApplied.20.044013,Wang2025IPA}. The security implications of these vulnerabilities under CV-QKD operating conditions remain insufficiently established experimentally. This limits the assessment of their practical significance and the measures needed to address them. Experimental investigations in CV-QKD systems, supported by security analyses based on measured results, are therefore needed to assess these vulnerabilities beyond their effects on individual components.

In this work, we investigate a reversible photorefractive attack in an LLO CV-QKD system and assess its security implications through experiments and analysis. Moderate blue-light illumination reduces the modulation variance of a lithium-niobate IQ modulator (IQM) to a stable value without an appreciable increase in estimated excess noise. The variance gradually recovers after illumination is removed. We elucidate the physical mechanisms underlying this response and analyze an attack scenario that exploits its reversibility to compromise modulation calibration. After the modulation variance recovers, analysis of receiver measurements using the calibration obtained during illumination reveals overestimated channel transmittance and underestimated excess noise. Key rate calculations based on measured parameters yield a positive estimated rate when the correctly calibrated rate is nonpositive. Recovery can also make the preceding optical manipulation more difficult to identify through later inspection. To our knowledge, this work is the first to experimentally reveal the security implications of a reversible photorefractive response in a CV-QKD system, demonstrating that its recovery can be exploited to compromise modulation calibration. These findings highlight reversible changes in quantum transmitters as a practical security concern, motivating security analyses and protection strategies that account for both optical perturbations and subsequent recovery.


\section{Results}
\label{sec:result}

\subsection{LLO CV-QKD system and photorefractive attack configuration}

\begin{figure*}[t]
\centering
\includegraphics[width=\textwidth]{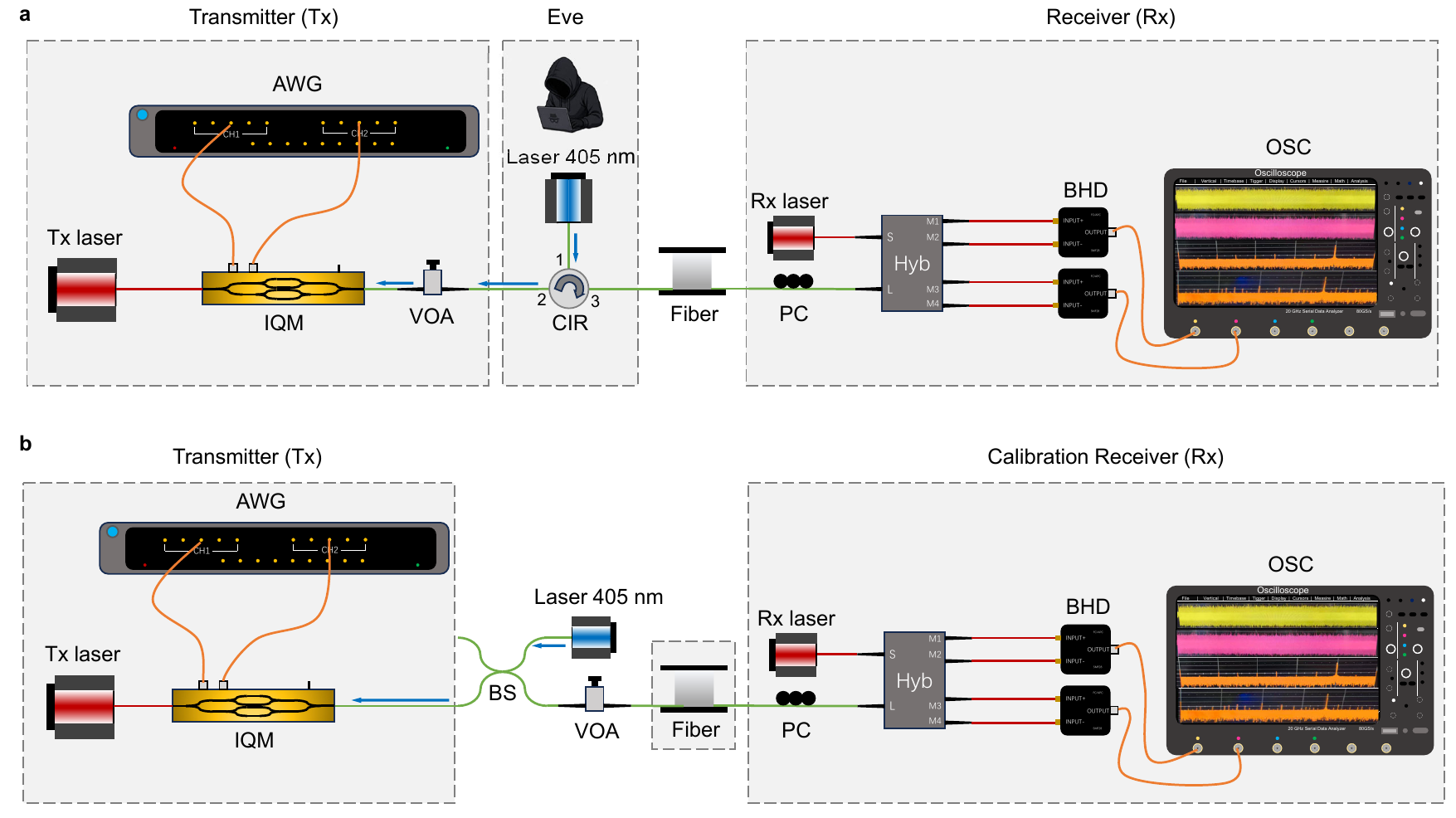}
\caption{
Schematic of the LLO CV-QKD system under photorefractive attack. (a) The 1550-nm optical carrier generated by the Tx laser is modulated by an IQM, where electrical modulation signals generated by an AWG are encoded onto the orthogonal quadratures of the optical field. The modulated signal light is subsequently attenuated to the quantum signal level by a VOA and transmitted through the quantum channel. After passing through the quantum channel, whose attenuation was measured to be approximately \SI{10.2}{dB}, the signal light reaches the Rx, where its polarization is compensated by a PC. The signal light is then combined with the local oscillator generated by the Rx laser through a 90$^\circ$ optical hybrid, and the interference outputs are detected by BHDs. The measurement results are recorded by an OSC. In the attack scenario, Eve would inject blue light into the Tx through the quantum channel to induce the photorefractive effect in the IQM. (b) To deliver sufficient blue-light power to the IQM with the available laser, the 405-nm light is introduced through a 50:50 BS between the IQM and the VOA, thereby bypassing the VOA. The resulting modulation changes are measured using a calibration receiver connected to the Tx. 
}
\label{Fig:LLO_CV_setup}
\end{figure*}

We constructed an LLO CV-QKD system to investigate the photorefractive attack on the transmitter. \Cref{Fig:LLO_CV_setup}(a) shows the transmitter (Tx), quantum channel, receiver (Rx), as well as Eve's blue-light injection through the quantum channel. At the Tx, the optical carrier is generated by a narrow-linewidth continuous-wave laser (Tx laser) with a center wavelength of \SI{1550.12}{\nano\meter} and is subsequently modulated by an IQM. An 8-bit arbitrary waveform generator (AWG) operating at a sampling rate of \SI{16}{GS/s} generates the electrical driving signals used to modulate the continuous-wave optical carrier through the IQM. The AWG output consists of a pilot signal and a quantum signal generated through frequency-division multiplexing. The quantum signal adopts a pulse-shaped Gaussian-modulated coherent-state (GMCS) format using a root-raised-cosine (RRC) filter and has a symbol rate of \SI{800}{MBaud}. It is subsequently frequency-shifted to an intermediate frequency centered at \SI{500}{MHz}. The pilot tone is located at \SI{-1}{GHz}, and the amplitude ratio between the pilot and quantum signals is fixed at $20$. After modulation, a VOA attenuates the optical signal to the power level required for CV-QKD before it enters the quantum channel. The VOA was adjusted to set the initial modulation variance to approximately $V_A=6.2$ shot-noise units (SNU). Its attenuation setting was then kept fixed throughout the subsequent measurements.

The signal light reaches the Rx after passing through the quantum channel, whose attenuation was measured to be approximately \SI{10.2}{dB}. A polarization controller (PC) is first used to align the polarization state of the incoming signal. The signal is then combined with the local oscillator generated by the Rx laser, which is identical to the Tx laser, using a 90$^\circ$ optical hybrid. Inside the optical hybrid, both the signal and LO are split into two paths, while the two LO branches maintain a relative phase difference of $\pi/2$. The resulting four optical outputs are detected by balanced homodyne detectors (BHDs) to perform heterodyne detection, and the corresponding electrical signals are acquired by an oscilloscope (OSC). The RF output bandwidth of each BHD is \SI{1.6}{GHz}, while the OSC operates at a sampling rate of \SI{40}{GS/s}. In the attack scenario shown in~\cref{Fig:LLO_CV_setup}(a), Eve would inject 405-nm blue light through the quantum channel and circulator toward the transmitter to induce a photorefractive response in the IQM.

After completing the optical and electrical connections, the shot noise and electronic noise of the CV-QKD system were calibrated. All modulation variances reported in this work are normalized to the SNU. To calibrate the electronic noise, only the BHDs and OSC were initially enabled while the local oscillator was blocked. Under this condition, the measured variance corresponds to the electronic noise, which was $4.37\times10^{-4}$. The Rx laser was then turned on, and the measured variance increased to $6.77\times10^{-3}$, corresponding to the sum of the shot noise and electronic noise. Consequently, the shot-noise variance $N_0$ was determined to be $6.33\times10^{-3}$, yielding an electronic noise level of $v_{\mathrm{el}}=0.069~\mathrm{SNU}$. The shot-noise calibration was repeated multiple times throughout the experiments. Owing to the excellent stability of the laser sources and optical setup, the measured shot-noise values remained nearly unchanged. The channel transmittance $T$ and excess noise $\xi$ were subsequently estimated from the correlated Gaussian variables prepared at the Tx and measured at the Rx. Details of the channel-parameter-estimation procedure are provided in Appendix~\ref{app:channel_parameter_estimation}. These calibrations and measurements provide the experimental basis for the subsequent security analysis.

\subsection{Photorefractive attack on the IQM}
\label{experiment_procedure}

The objective of this experiment is to characterize the photorefractive response of the IQM before assessing its security implications in the LLO CV-QKD system. Since the IQM is located in the Tx, the quantum channel was initially removed, and the Tx was directly connected to a calibration receiver. This configuration eliminates the influence of channel attenuation and enables the modulation characteristics of the IQM to be directly evaluated under blue-light illumination. The experimental configuration is illustrated in~\cref{Fig:LLO_CV_setup}(b). After characterizing the modulation changes, the quantum channel was reconnected to measure the relevant system parameters under blue-light illumination.

Blue light entering the Tx through the quantum channel must pass through the VOA before reaching the IQM, as shown in~\cref{Fig:LLO_CV_setup}(a). Our measurements show that the VOA introduces approximately $57$--$58~\mathrm{dB}$ attenuation at \SI{405}{\nano\meter}. The commercial 1550-nm optical circulator in~\cref{Fig:LLO_CV_setup}(a) also has substantial loss at this wavelength, with measured attenuation lower bounds ranging from $71.8$ to $75.4~\mathrm{dB}$. With the available blue laser, these losses would leave insufficient optical power at the IQM for the experiment. We therefore introduced the 405-nm light through a 50:50 beam splitter (BS) between the IQM and the VOA, as shown in~\cref{Fig:LLO_CV_setup}(b). This arrangement bypassed the VOA for the blue light and allowed us to control the power incident on the IQM. The measured attenuation of the BS was approximately $3.6$--$4.3~\mathrm{dB}$. Based on the VOA attenuation alone, delivering approximately $5~\mu\mathrm{W}$ of blue light to the IQM through the VOA would require several watts at its input. A simplified estimate gives a nominal continuous-wave fiber damage threshold of approximately \SI{6.54}{W} at \SI{405}{\nano\meter}. These estimates suggest that Eve could inject blue light through the quantum channel to affect the IQM, provided that the injection path has sufficiently low loss at 405 nm. Detailed component measurements and the threshold estimate are provided in Appendix~\ref{App:405_setup_LIDT}. The experimental procedure is summarized as follows.

\textbf{Step 1. Initial modulation variance calibration.}
After calibrating the electronic noise and shot noise, the Tx laser was turned on, and the IQM was driven by the electrical waveform generated by the AWG. The VOA was adjusted such that, after coherent detection and digital signal processing (DSP) at the calibration receiver, the initial modulation variance was approximately $V_A=6.2~\mathrm{SNU}$. The VOA setting was kept fixed during the subsequent measurements. The detailed DSP procedure is described in Appendix~\ref{App:DSP}. The initial modulation variance was determined from three independent measurements.

\textbf{Step 2. Blue-light injection.}
The blue laser was then switched on with an initial optical power of $1~\mu\mathrm{W}$ incident on the IQM. Under continuous blue-light illumination, the modulation variance was monitored as a function of time. Measurements were recorded at $1~\mathrm{s}$, $10~\mathrm{s}$, $20~\mathrm{s}$, $30~\mathrm{s}$, $40~\mathrm{s}$, $50~\mathrm{s}$, $1~\mathrm{min}$, $1.5~\mathrm{min}$, $2~\mathrm{min}$, $3~\mathrm{min}$, $4~\mathrm{min}$, $5~\mathrm{min}$, $6~\mathrm{min}$, $7~\mathrm{min}$, $8~\mathrm{min}$, $9~\mathrm{min}$, $11~\mathrm{min}$, $13~\mathrm{min}$, $15~\mathrm{min}$, $18~\mathrm{min}$, $21~\mathrm{min}$, $24~\mathrm{min}$, $29~\mathrm{min}$, $34~\mathrm{min}$, $39~\mathrm{min}$, $47~\mathrm{min}$, $55~\mathrm{min}$, $63~\mathrm{min}$, $76~\mathrm{min}$, $89~\mathrm{min}$, $105~\mathrm{min}$, $125~\mathrm{min}$, $150~\mathrm{min}$, and $180~\mathrm{min}$.

\textbf{Step 3. Recovery and post-illumination measurement.}
After \SI{180}{min} of illumination, the blue laser was switched off. The recovery behavior of the IQM was then monitored by measuring the modulation variance at $0~\mathrm{s}$, $1~\mathrm{min}$, $2~\mathrm{min}$, $3~\mathrm{min}$, $4~\mathrm{min}$, $5~\mathrm{min}$, $6~\mathrm{min}$, $7~\mathrm{min}$, $8~\mathrm{min}$, $9~\mathrm{min}$, $10~\mathrm{min}$, $15~\mathrm{min}$, $30~\mathrm{min}$, $1~\mathrm{h}$, $2~\mathrm{h}$, and $8~\mathrm{h}$. After the IQM had recovered and reached a stable operating state, its modulation variance was determined from three independent measurements.

\textbf{Step 4. Optical-power escalation.}
If the IQM remained operational throughout the blue-light illumination and continued to perform reliable information encoding, the incident blue-light power was increased by $1~\mu\mathrm{W}$, and Steps 2 and 3 were repeated. This process continued until the IQM exhibited severe performance degradation and could no longer maintain stable modulation, at which point the power sweep was stopped.

\textbf{Step 5. System-level parameter measurement.}
Finally, the quantum channel was reconnected, and the experimental procedure described above was repeated under the same illumination and measurement conditions. The channel transmittance and excess noise were estimated from the recorded data. The modulation variances measured during stable illumination and after recovery formed the basis for the subsequent security analysis of the attack scenario.

\subsection{Photorefractive response of the IQM}

\begin{figure*}[t]
\centering
\includegraphics[width=\textwidth]{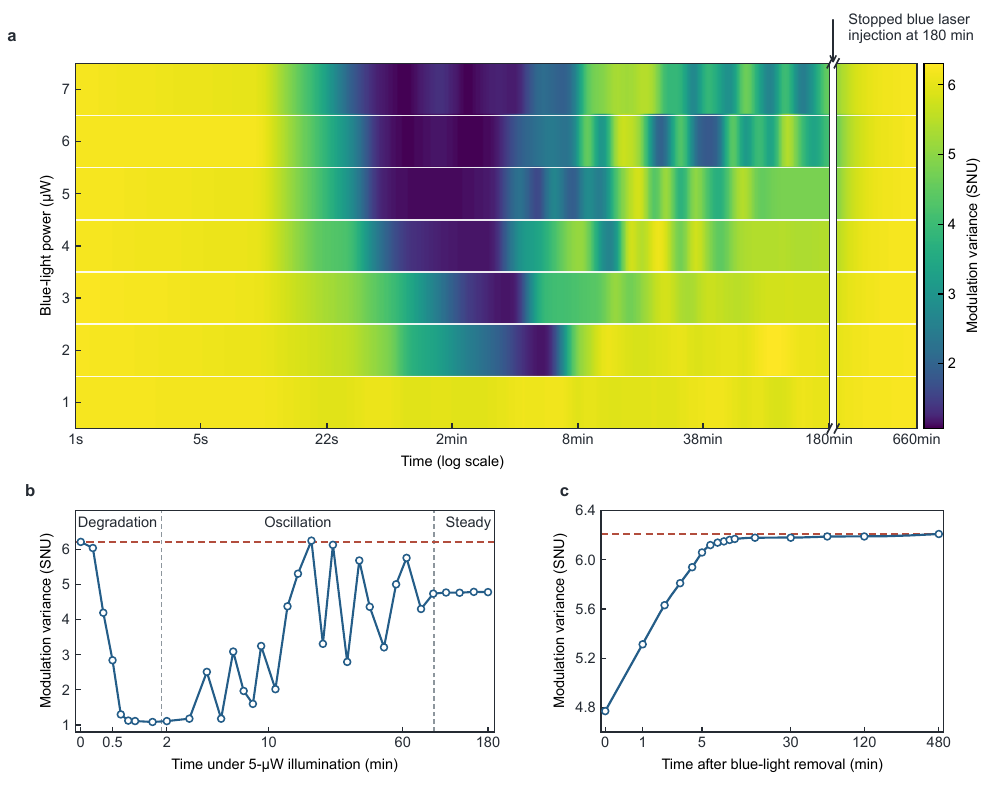}
\caption{
Photorefractive response of the IQM under blue-light illumination and after its removal. (a) Temporal evolution of the measured modulation variance at incident blue-light powers from $1$ to $7~\mu\mathrm{W}$. The color scale gives the modulation variance in SNU. Blue-light illumination is stopped after \SI{180}{min}. (b) Measured modulation variance during $5~\mu\mathrm{W}$ illumination, showing the degradation, oscillation, and steady stages. The vertical dashed lines indicate approximate stage boundaries. (c) Recovery of the modulation variance after removal of the $5~\mu\mathrm{W}$ illumination. In (b) and (c), open circles indicate measured values, solid lines guide the eye, and horizontal dashed lines mark the pre-illumination variance.
}
\label{Fig:main_res}
\end{figure*}

The experimental results obtained from Steps 2--5 are presented in this section. We first examine the temporal evolution of the IQM modulation variance under blue-light illumination, followed by its recovery after the illumination is removed. The results are visualized as a heat map in~\cref{Fig:main_res}(a). The horizontal axis uses a logarithmic scale to show the rapid changes after the onset of illumination. The vertical axis denotes the blue-light power incident on the IQM, while the color scale represents the measured modulation variance. Before the blue-light illumination is removed, the evolution of the modulation variance can be divided into three distinct stages---a degradation stage, an oscillation stage, and a steady stage. The onset time and duration of each stage depend on the incident blue-light power. After the illumination is removed, the modulation variance gradually recovers toward its pre-illumination value. The main experimental results are summarized as follows.

\textbf{Degradation stage.}
Immediately after the onset of blue-light illumination, the modulation variance decreases rapidly, which is defined as the degradation stage. When the incident power is only $1~\mu\mathrm{W}$, the reduction in modulation variance is relatively small. However, for incident powers of $2~\mu\mathrm{W}$ and above, the modulation variance decreases significantly. Increasing the blue-light power also causes the degradation stage to occur earlier, accelerates the decrease in modulation variance, and leads to a lower minimum modulation variance.

\textbf{Oscillation and steady stages.}
Following the rapid degradation, the modulation variance gradually increases but exhibits noticeable fluctuations, which is defined as the oscillation stage. During this stage, the variance does not monotonically return toward its initial value but fluctuates under continuous blue-light illumination. If the modulation variance eventually converges to a stable value, the subsequent period is referred to as the steady stage. For incident powers no higher than $5~\mu\mathrm{W}$, the variance gradually reaches a stable value after the oscillation stage, indicating that the IQM reaches a new stable operating condition. As the incident blue-light power increases, the oscillation stage becomes longer. The stabilized modulation variance also decreases. The $5~\mu\mathrm{W}$ trace in~\cref{Fig:main_res}(b) shows these stages in detail. In contrast, at $6~\mu\mathrm{W}$ and $7~\mu\mathrm{W}$, the modulation variance continues to fluctuate throughout the \SI{180}{min} illumination period without reaching a stable state.

\textbf{Recovery after blue-light removal.}
After \SI{180}{min} of continuous illumination, the blue laser was switched off, and the modulation variance gradually recovered toward its pre-illumination value. After approximately \SI{8}{h}, the variance had nearly returned to its original level. \Cref{tab:modulation_variance} compares the variance before illumination, after stabilization under illumination where a stable state was reached, and after recovery. The recovery process at $5~\mu\mathrm{W}$ is shown in~\cref{Fig:main_res}(c). It exhibits two distinct timescales. The modulation variance increases rapidly and approaches its original value within approximately \SI{6}{min}, followed by a slower relaxation toward the pre-illumination level over several hours.

\begin{table}[t]
\caption{IQM modulation variance before blue-light illumination, after stabilization under illumination, and after recovery. All values are in SNU. A dash indicates that the variance did not stabilize during the \SI{180}{min} illumination period.}
\label{tab:modulation_variance}
\centering
\begin{tabular}{cccc}
\toprule
\shortstack{Blue-light\\power}
& \shortstack{Before\\illumination}
& \shortstack{After\\stabilization}
& \shortstack{After\\recovery} \\
\midrule
$1~\mu\mathrm{W}$ & 6.17 & 6.05 & 6.18 \\
$2~\mu\mathrm{W}$ & 6.22 & 5.99 & 6.22 \\
$3~\mu\mathrm{W}$ & 6.21 & 5.77 & 6.25 \\
$4~\mu\mathrm{W}$ & 6.20 & 5.41 & 6.20 \\
$5~\mu\mathrm{W}$ & 6.21 & 4.78 & 6.21 \\
$6~\mu\mathrm{W}$ & 6.19 & \textemdash & 6.24 \\
$7~\mu\mathrm{W}$ & 6.18 & \textemdash & 6.16 \\
\bottomrule
\end{tabular}
\end{table}

\begin{figure*}[t]
\centering
\includegraphics[width=\textwidth]{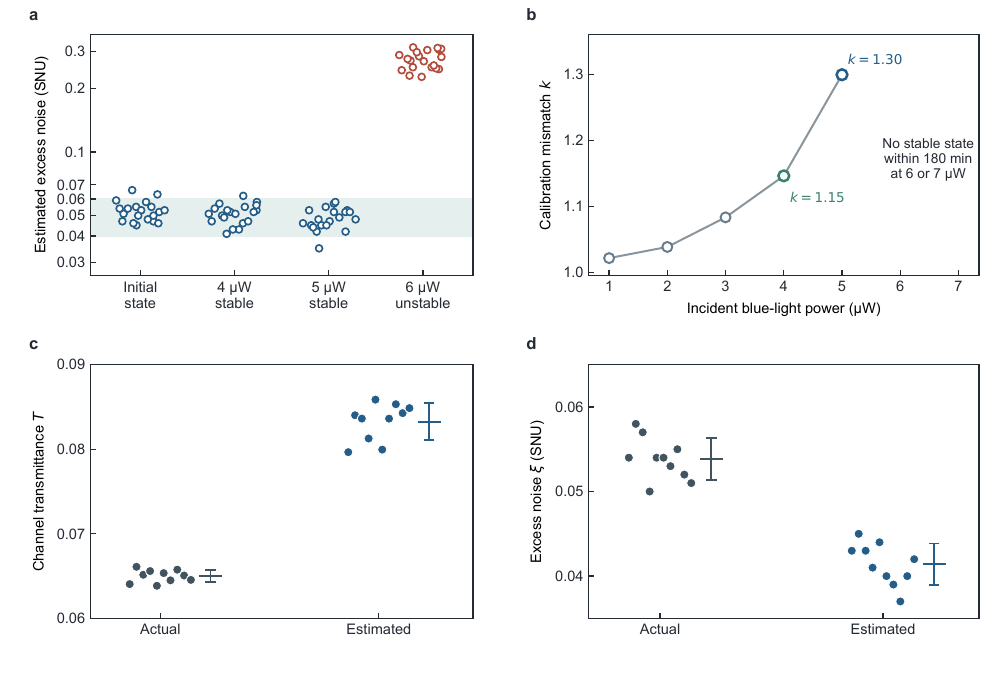}
\caption{
Experimental measurements relevant to the photorefractive attack. (a) Excess-noise estimates from 20 independent acquisitions per condition before illumination, after stabilization under $4~\mu\mathrm{W}$ or $5~\mu\mathrm{W}$ illumination, and during $6~\mu\mathrm{W}$ illumination. The modulation variance remains unstable in the last condition. The shaded band marks $0.04$--$0.06~\mathrm{SNU}$. (b) Calibration mismatch $k=V_A'/V_A$ calculated from the stabilized and recovered modulation variances in~\cref{tab:modulation_variance}. No stable illuminated state was reached at $6~\mu\mathrm{W}$ or $7~\mu\mathrm{W}$. (c,d) Channel transmittance and excess noise estimated after recovery from $5~\mu\mathrm{W}$ illumination. Actual denotes values calculated using the modulation variance after recovery, while Estimated denotes values calculated using the variance measured during illumination. Each group contains ten independent acquisitions. Horizontal markers show the group means, and error bars indicate one standard deviation.
}
\label{Fig:attack_parameters}
\end{figure*}

The preceding measurements show that the IQM can reach a stable operating condition with reduced modulation variance under blue-light illumination. The variance then recovers after the illumination is removed. To assess the effect of illumination on excess noise, we analyzed receiver data obtained with the quantum channel connected, as described in Step 5. Twenty independent data sets were acquired under each measurement condition and processed using the DSP and channel-parameter-estimation procedures. Measurements were made before illumination, after stabilization under $4~\mu\mathrm{W}$ or $5~\mu\mathrm{W}$ illumination, and during $6~\mu\mathrm{W}$ illumination. In the last condition, the modulation variance continued to fluctuate without reaching a steady state. The resulting estimates are shown in~\cref{Fig:attack_parameters}(a). Across the first three conditions, 56 of the 60 estimates lie within $0.04$--$0.06~\mathrm{SNU}$. Thus, the reduced modulation variance at $4~\mu\mathrm{W}$ and $5~\mu\mathrm{W}$ is not accompanied by an appreciable increase in estimated excess noise. During $6~\mu\mathrm{W}$ illumination, the estimates rise to $0.22$--$0.31~\mathrm{SNU}$. Such an increase could alert Alice and Bob to the disturbance. We therefore use the stable $4~\mu\mathrm{W}$ and $5~\mu\mathrm{W}$ conditions in the following security analysis.

\subsection{Attack strategy and security analysis}

The experimental results show that the IQM can maintain stable modulation under moderate blue-light illumination, although its modulation variance decreases. After the illumination is removed, the variance gradually returns toward its initial value. These observations motivate the following attack scenario. Eve would illuminate the IQM while Alice calibrates the transmitter modulation. For example, an incident blue-light power of $5~\mu\mathrm{W}$ produces a stable modulation variance of $V_A=4.78~\mathrm{SNU}$. Once the calibration is complete, Eve would remove the illumination. After the variance has recovered, Alice and Bob would collect data for channel-parameter estimation and key extraction. Alice would still use the modulation variance determined under illumination. This mismatch could cause Alice and Bob to overestimate the secret key rate. If they determine the final key length from that estimate, privacy amplification may fail to remove Eve's potential information about the key, thereby threatening the practical security of CV-QKD. We next quantify the resulting biases in the estimated channel parameters and secret key rate.

Let $x_A$ denote Alice's modulation variable according to the calibration performed under illumination, with variance $V_A$. After the IQM recovers, the actual encoded variable becomes $x_A'=\sqrt{k}x_A$, where $k=V_A'/V_A$ and $V_A'$ is the recovered modulation variance. Assuming that the physical channel transmittance $T$ and excess noise $\xi$ remain unchanged, Alice and Bob estimate the channel from data collected after recovery using $x_A$. This gives
\[
T_{\mathrm{est}}=kT,
\qquad
\xi_{\mathrm{est}}=\frac{\xi}{k}.
\]
The derivation is provided in Appendix~\ref{app:channel_parameter_estimation}, following Ref.~\cite{PhysRevA.100.012313}. The channel consequently appears more transmissive and less noisy than it is. Alice and Bob calculate the estimated key rate using $(V_A,T_{\mathrm{est}},\xi_{\mathrm{est}})$, whereas the rate calculated using $(V_A',T,\xi)$ is labeled the actual rate below.

The measured modulation variances give the mismatch ratios shown in~\cref{Fig:attack_parameters}(b). Under $4~\mu\mathrm{W}$ illumination, the variance decreases to $5.41~\mathrm{SNU}$ and returns to $6.20~\mathrm{SNU}$ after recovery, giving $k=1.15$. The corresponding values for $5~\mu\mathrm{W}$ are $4.78$ and $6.21~\mathrm{SNU}$, giving $k=1.30$. We further examine the parameter bias for the $5~\mu\mathrm{W}$ condition using receiver data acquired after recovery. Ten independent acquisitions were analyzed using the recovered modulation variance $V_A'$, and another ten were analyzed using the earlier value $V_A$. These analyses provide the actual and estimated channel parameters, respectively. As shown in~\cref{Fig:attack_parameters}(c) and~\cref{Fig:attack_parameters}(d), the mean actual values are $T=0.065$~\cite{note:effective_loss} and $\xi=0.0538~\mathrm{SNU}$. The mean estimated values are $T_{\mathrm{est}}=0.083$ and $\xi_{\mathrm{est}}=0.0414~\mathrm{SNU}$. The resulting ratios are $T_{\mathrm{est}}/T=1.28$ and $\xi/\xi_{\mathrm{est}}=1.30$, in close agreement with the mismatch ratio $k=1.30$ obtained from the measured modulation variances.

\begin{figure*}[t]
\centering
\includegraphics[width=\textwidth]{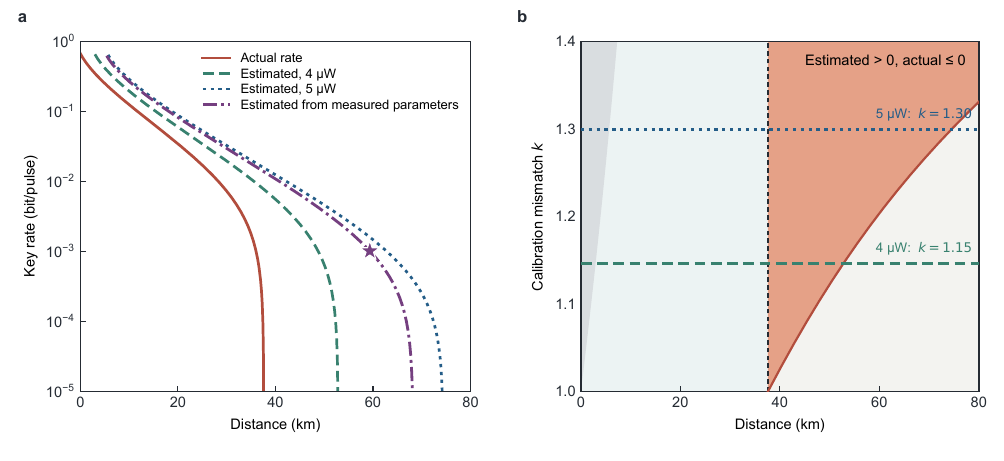}
\caption{
Asymptotic key-rate analysis of the IQM photorefractive attack. (a) The solid line shows the actual rate calculated using the recovered modulation variance. The dashed and dotted lines show the rates estimated using the modulation variances measured under $4~\mu\mathrm{W}$ and $5~\mu\mathrm{W}$ illumination, respectively. The purple line uses the estimated parameters obtained from receiver measurements for the $5~\mu\mathrm{W}$ condition. The star marks the estimated rate under the measured system conditions. The estimated curves begin at nonzero distances because their apparent channel transmittance would exceed unity at shorter distances. (b) Key-rate outcomes at different calibration mismatch ratios $k$ and transmission distances. The orange region indicates a positive estimated rate when the actual rate is nonpositive. The horizontal dashed and dotted lines mark the measured $k$ values for $4~\mu\mathrm{W}$ and $5~\mu\mathrm{W}$, respectively. The gray region corresponds to an apparent channel transmittance greater than unity.
}
\label{Fig:attack_keyrate}
\end{figure*}

Using the measured modulation variances and the formulation in Appendix~\ref{App:keyrate_sim}, we calculate the asymptotic key rates shown in~\cref{Fig:attack_keyrate}(a). The solid curve represents the actual rate after recovery. The dashed and dotted curves represent the rates estimated using the modulation variances measured under $4~\mu\mathrm{W}$ and $5~\mu\mathrm{W}$ illumination, respectively. For these three curves, we use $\xi=0.05~\mathrm{SNU}$ as a representative physical excess noise based on the measurements reported above. The purple curve is based on the estimated channel parameters for the $5~\mu\mathrm{W}$ condition. All curves use the measured electronic noise $v_{\mathrm{el}}=0.069~\mathrm{SNU}$, detector efficiency $\eta=0.526$, reconciliation efficiency $\beta=0.89$, and fiber attenuation coefficient $\alpha=0.2~\mathrm{dB/km}$.

As shown in~\cref{Fig:attack_keyrate}(a), the actual rate becomes nonpositive beyond approximately \SI{38}{km}. At \SI{40}{km}, however, the modulation variances measured under $4~\mu\mathrm{W}$ and $5~\mu\mathrm{W}$ illumination give estimated rates of approximately $5.6\times10^{-3}$ and $1.25\times10^{-2}$ bit/pulse, respectively. Under the measured system conditions, the purple star indicates that Alice and Bob would estimate a key rate of $1.02\times10^{-3}$ bit/pulse, whereas the actual rate is nonpositive. As shown in~\cref{Fig:attack_keyrate}(b), the range in which Alice and Bob incorrectly estimate a positive key rate grows with the calibration mismatch. For the measured ratios $k=1.15$ and $k=1.30$, this range extends from approximately \SI{38}{km} to \SI{53}{km} and \SI{74}{km}, respectively. These results show that the reversible photorefractive perturbation can threaten CV-QKD security when Alice and Bob continue to use the transmitter calibration obtained under illumination after the modulation variance has recovered.


\section{Discussion}
\label{sec:discussion}

The results demonstrate that stable modulation and device recovery do not ensure the continued validity of transmitter calibration. At suitable incident powers, the modulation variance changes without an appreciable increase in estimated excess noise, indicating that a low noise estimate alone cannot establish consistency between the calibration and the transmitted states. After the modulation variance recovers, analysis of receiver measurements using the calibration obtained during illumination confirms overestimated channel transmittance and underestimated excess noise. The measured biases are consistent with the calibration mismatch predicted from the change in modulation variance, supporting the proposed attack mechanism. Asymptotic key rate calculations based on measured parameters further show that this mismatch can produce a positive estimated rate when the correctly calibrated rate is nonpositive. If Alice and Bob determine the final key length from the overestimated rate, privacy amplification may be insufficient to remove Eve's potential information. Recovery also makes the earlier disturbance more difficult to identify through later inspection. The resulting risk therefore extends beyond the period of illumination when the affected calibration remains in use.

Mitigating this attack requires verification that the transmitter calibration remains consistent with the states used for parameter estimation and key generation. Alice could monitor the modulation variance during operation and use changes in the IQM bias as an additional warning signal. Detected changes should trigger a calibration check and, when necessary, recalibration. Data for which the modulation calibration cannot be verified should be excluded from key generation. Spectral filtering, optical isolation, and optical power monitoring at possible entry points could suppress or reveal injected light, but their effectiveness requires verification at the attack wavelength. Our experiment used an illumination path that bypassed the VOA to deliver sufficient blue-light power to the IQM, so injection through the full quantum channel path remains to be demonstrated experimentally. The simulations compare stable operating states without modeling the continuous change in modulation variance during recovery. Future work could examine external injection through the quantum channel and parameter estimation during recovery, followed by an evaluation of monitoring and recalibration measures within a finite-key analysis. These investigations would support the development of security analyses and protection strategies that account for both optical perturbations and the subsequent recovery of quantum transmitters.


\section{Methods}
\label{sec:method}

As discussed in~\cref{sec:result}, the IQM exhibits three distinct stages under blue-light illumination, namely degradation, oscillation, and steady operation. Under stable $4$ and $5~\mu\mathrm{W}$ illumination, the measured modulation variance decreases while the excess-noise estimates remain close to their pre-illumination levels. In this section, we explain the observed modulation response in terms of photoinduced phase drift, ABC compensation, and optical attenuation.

\subsection{Operating principle and initialization of the IQ modulator}

An IQM comprises two Mach--Zehnder modulators (MZMs) that encode the in-phase (I) and quadrature (Q) components of the optical field. After independent modulation, their optical fields are combined with a relative phase of $\phi=\pi/2$. Neglecting additional device loss, the output optical field and power can be expressed as
\begin{equation}
E_{\mathrm{out}}
=
\frac{1}{2}E_{\mathrm{in}}
\left(
I_I+I_Qe^{j\phi}
\right),
\end{equation}
\begin{equation}
\begin{aligned}[c]
P_{\mathrm{out}}
&=
|E_{\mathrm{out}}|^2\\
&=
\frac{1}{4}P_{\mathrm{in}}
\left(
I_I^2+I_Q^2+2I_I I_Q\cos\phi
\right).
\end{aligned}
\end{equation}
Here,
\begin{equation}
\begin{aligned}[c]
I_I
&=
\cos\left(
\frac{\pi}{2}
\frac{V_I}{V_\pi}
\right),\\
I_Q
&=
\cos\left(
\frac{\pi}{2}
\frac{V_Q}{V_\pi}
\right)
\end{aligned}
\end{equation}
are the normalized optical-field transmission functions of the I- and Q-branch MZMs, respectively. $V_\pi$ denotes the half-wave voltage, while $V_I$ and $V_Q$ are the effective driving voltages of the two branches. The input optical power is defined as
\begin{equation}
P_{\mathrm{in}}=|E_{\mathrm{in}}|^2.
\end{equation}

For each branch, the driving voltage consists of a high-speed modulation signal and a DC bias voltage,
\begin{equation}
\begin{aligned}[c]
V_I
&=
V_{\mathrm{dataI}}+V_{\mathrm{biasI}},\\
V_Q
&=
V_{\mathrm{dataQ}}+V_{\mathrm{biasQ}},
\end{aligned}
\end{equation}
where $V_{\mathrm{dataI}}$ and $V_{\mathrm{dataQ}}$ denote the modulation signals applied to the I and Q branches. The DC bias voltages $V_{\mathrm{biasI}}$ and $V_{\mathrm{biasQ}}$ are controlled by the automatic bias controller (ABC). After powering on, the IQM enters an initialization stage in which the continuous-wave laser is activated without the random modulation signals. Thus,
\begin{equation}
V_{\mathrm{dataI}}=V_{\mathrm{dataQ}}=0,
\end{equation}
and the effective driving voltages reduce to
\begin{equation}
\begin{aligned}[c]
V_I
&=
V_{\mathrm{biasI}},\\
V_Q
&=
V_{\mathrm{biasQ}}.
\end{aligned}
\end{equation}

During initialization, the ABC monitors the output optical power and adjusts the DC bias voltages toward the predefined operating point. In this work, both MZMs are biased near their minimum-transmission points. In the idealized model, their target bias voltages satisfy
\begin{equation}
\begin{aligned}[c]
V_{\mathrm{biasI}}
&\rightarrow V_\pi,\\
V_{\mathrm{biasQ}}
&\rightarrow V_\pi.
\end{aligned}
\end{equation}
The ideal output power consequently approaches zero. In a practical device, finite extinction ratio, waveguide loss, and fabrication imperfections leave a nonzero minimum output power. The ABC maintains the IQM near this initial operating condition by adjusting the DC biases rather than the high-speed modulation signals. When an external perturbation shifts the MZM operating points, the ABC adjusts the DC biases to restore stable operation, provided that the shift remains within its tracking range.

\subsection{Physical mechanisms of blue-light-induced perturbations in the IQ modulator}
\label{Equivalent_Physical_Model}

\subsubsection{Space-charge field and refractive-index variation}

When 405-nm blue light enters an operating IQM, it induces a photorefractive response in the $\mathrm{LiNbO_3}$ material. Photoexcitation, carrier transport, and trapping establish a space-charge field that changes the refractive index~\cite{PhysRevApplied.19.054052,PhysRevApplied.20.044013}. Only the differential phase shift between the two interferometer arms changes an MZM operating point. We therefore use $E_{\mathrm{sc}}$ to denote the differential component of the photoinduced space-charge field and $\Delta n$ to denote the corresponding difference in effective refractive-index change. As a first-order approximation within the investigated illumination range, we take $E_{\mathrm{sc}}$ to be proportional to the blue-light power incident on the IQM,
\begin{equation}
E_{\mathrm{sc}}\propto P_b,
\end{equation}
where $P_b$ denotes that incident power.

According to the Pockels effect, the effective field produces a differential refractive-index change,
\begin{equation}
\Delta n
=
-\frac{1}{2}n^3r_{33}E_{\mathrm{sc}},
\end{equation}
where $n$ is the relevant refractive index and $r_{33}$ is the electro-optic coefficient used for the configuration considered here. Under the stated approximation, the index change follows
\begin{equation}
\Delta n\propto P_b.
\end{equation}
This differential index change introduces a relative phase shift between the two MZM arms,
\begin{equation}
\Delta\phi_{\mathrm{ph}}
=
\frac{2\pi}{\lambda}\Delta n L,
\end{equation}
where $\lambda$ is the wavelength of the modulated 1550-nm light and $L$ is the effective interaction length. Consequently, within the same approximation,
\begin{equation}
\Delta\phi_{\mathrm{ph}}\propto P_b.
\end{equation}

\subsubsection{Equivalent bias drift}

In an MZM, the applied voltage changes the relative optical phase between its two arms through the electro-optic effect. The voltage-induced phase shift is
\begin{equation}
\Delta\phi_{\mathrm{EO}}
=
\pi\frac{V}{V_\pi}.
\end{equation}
Because the voltage-induced and photoinduced phase shifts act on the same relative phase, the photorefractive perturbation corresponds to an equivalent displacement of the MZM bias point. For one MZM branch, the total phase variation under illumination is
\begin{equation}
\Delta\phi_{\mathrm{tot}}
=
\pi\frac{V}{V_\pi}
+
\Delta\phi_{\mathrm{ph}}.
\end{equation}
Defining the equivalent bias shift as
\begin{equation}
\Delta V^{(\mathrm{ph})}
=
\frac{V_\pi}{\pi}
\Delta\phi_{\mathrm{ph}}
\end{equation}
allows the total phase variation to be written as
\begin{equation}
\Delta\phi_{\mathrm{tot}}
=
\pi
\frac{V+\Delta V^{(\mathrm{ph})}}{V_\pi}.
\end{equation}
The term $\Delta V^{(\mathrm{ph})}$ describes an optical perturbation in the MZM transfer function; it does not imply that blue-light illumination generates an electrical voltage on the electrodes.

For the two MZM branches of an IQM, the equivalent bias shifts are
\begin{equation}
\begin{aligned}[c]
\Delta V_I^{(\mathrm{ph})}
&=
\frac{V_{\pi,I}}{\pi}
\Delta\phi_{\mathrm{ph},I},\\
\Delta V_Q^{(\mathrm{ph})}
&=
\frac{V_{\pi,Q}}{\pi}
\Delta\phi_{\mathrm{ph},Q}.
\end{aligned}
\end{equation}
The 405-nm light need not propagate identically through the two branches because the splitter response and waveguide modes can depend on wavelength. Their photoinduced phase shifts can consequently differ,
\begin{equation}
\Delta\phi_{\mathrm{ph},I}
\neq
\Delta\phi_{\mathrm{ph},Q}.
\end{equation}
The corresponding equivalent bias shifts can also differ,
\begin{equation}
\Delta V_I^{(\mathrm{ph})}
\neq
\Delta V_Q^{(\mathrm{ph})}.
\end{equation}

\subsubsection{Blue-light-induced increase in insertion loss}

In addition to bias drift, previous experiments have shown that blue-light illumination can increase the insertion loss of $\mathrm{LiNbO_3}$ intensity and phase modulators, with greater loss at higher blue-light power~\cite{PhysRevApplied.20.044013}. Motivated by these reports, we represent the additional loss in the I and Q branches by field-amplitude transmission coefficients $\eta_I(P_b)$ and $\eta_Q(P_b)$. Their squares give the corresponding power-transmission factors. We assume that $0<\eta_I,\eta_Q\leq1$ and that both coefficients decrease with $P_b$ over the range considered,
\begin{equation}
\frac{d\eta_I}{dP_b}<0,
\qquad
\frac{d\eta_Q}{dP_b}<0.
\label{eq:dandiao}
\end{equation}

Including the common splitting and combining factor, the contributions of the two branches to the IQM output field are
\begin{equation}
\begin{aligned}[c]
E_I
&=
\frac{1}{2}\eta_I(P_b)I_I E_{\mathrm{in}},\\
E_Q
&=
\frac{1}{2}\eta_Q(P_b)I_Q E_{\mathrm{in}}e^{j\phi}.
\end{aligned}
\end{equation}
Their sum gives the IQM output field,
\begin{equation}
E_{\mathrm{out}}
=
E_I+E_Q
=
\frac{1}{2}E_{\mathrm{in}}
\left(
\eta_I I_I
+
\eta_Q I_Q e^{j\phi}
\right).
\end{equation}
The corresponding output power is
\begin{equation}
P_{\mathrm{out}}
=
\frac{1}{4}P_{\mathrm{in}}
\left(
\eta_I^2I_I^2
+
\eta_Q^2I_Q^2
+
2\eta_I\eta_QI_I I_Q\cos\phi
\right).
\label{eq:p_out}
\end{equation}

\subsection{ABC compensation and stabilization under blue-light illumination}
\label{ABC_analysis}

When blue-light illumination begins, the photoinduced refractive-index change introduces an additional optical phase shift. This shift can be represented as an equivalent DC-bias drift in the MZM transfer function. The effective driving voltages of the two branches then become
\begin{equation}
\begin{aligned}[c]
V_I
&=
V_{\mathrm{biasI}}
+
V_{\mathrm{dataI}}
+
\Delta V_I^{(\mathrm{ph})},\\
V_Q
&=
V_{\mathrm{biasQ}}
+
V_{\mathrm{dataQ}}
+
\Delta V_Q^{(\mathrm{ph})}.
\end{aligned}
\end{equation}
At the onset of illumination, the photoinduced phase shift displaces the operating points from their initial settings. This displacement explains the initial decrease in modulation variance, while feedback adjustment provides a mechanism for the subsequent fluctuations.

We describe the feedback process using a simplified dither-based formulation of ABC operation~\cite{Li:17}. Low-frequency bias modulation allows the optical output to provide feedback error signals. For the $n$th averaging interval, these signals can be represented as
\begin{equation}
\begin{aligned}[c]
CO_I^{(n)}
&=
\frac{1}{\tau}
\int_{t_n}^{t_n+\tau}
P_{\mathrm{out}}(t)
\sin(\omega_1 t)\,dt,\\
CO_Q^{(n)}
&=
\frac{1}{\tau}
\int_{t_n}^{t_n+\tau}
P_{\mathrm{out}}(t)
\sin(\omega_2 t)\,dt.
\end{aligned}
\end{equation}
Here, $\tau$ is the averaging interval, while $\omega_1$ and $\omega_2$ are the dither angular frequencies used in this description. The signals $CO_I^{(n)}$ and $CO_Q^{(n)}$ indicate deviations of the respective branch bias points from their target operating points. A simplified negative-feedback update can be written as
\begin{equation}
\begin{aligned}[c]
V_{\mathrm{biasI}}^{(n+1)}
&=
V_{\mathrm{biasI}}^{(n)}
-
\mu_I CO_I^{(n)},\\
V_{\mathrm{biasQ}}^{(n+1)}
&=
V_{\mathrm{biasQ}}^{(n)}
-
\mu_Q CO_Q^{(n)},
\end{aligned}
\end{equation}
where $\mu_I$ and $\mu_Q$ are feedback gains for the two branches.

\begin{figure}[htbp]
\centering
\includegraphics[width=0.9\linewidth]{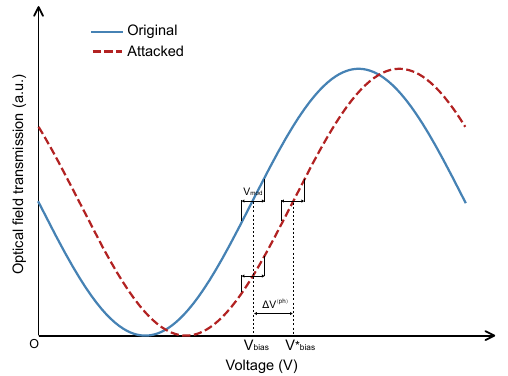}
\caption{
Bias-voltage evolution and optical-transmission characteristics of the IQM under the photorefractive effect. The blue curves represent the original operating condition, where the I- and Q-branch bias voltages are initialized at the target operating points. The modulation-voltage range is denoted by $V_{\mathrm{mod}}$. The red curves represent the condition after the photorefractive perturbation, where the equivalent bias shifts by $\Delta V^{(\mathrm{ph})}$. The ABC controller subsequently adjusts the DC bias voltage and relocks the modulator to a stable target operating condition, resulting in a new bias voltage $V_{\mathrm{bias}}^{*}$.
}
\label{Fig:V_pi}
\end{figure}

As the feedback loop converges, repeated bias adjustments reduce the error signals. At ideal convergence,
\begin{equation}
CO_I^{(n)}\rightarrow0,
\qquad
CO_Q^{(n)}\rightarrow0,
\end{equation}
and the bias voltages approach new steady-state values,
\begin{equation}
\begin{aligned}[c]
V_{\mathrm{biasI}}
&\rightarrow
V_{\mathrm{biasI}}^{*},\\
V_{\mathrm{biasQ}}
&\rightarrow
V_{\mathrm{biasQ}}^{*}.
\end{aligned}
\end{equation}
The quantities $V_{\mathrm{biasI}}^{*}$ and $V_{\mathrm{biasQ}}^{*}$ denote the steady-state bias voltages reached upon feedback convergence. Their evolution and the associated optical-transmission characteristics are illustrated schematically in~\cref{Fig:V_pi}.

ABC feedback compensates for photoinduced bias drift, explaining the transition from the oscillation stage to stable modulation. The measured modulation variance nevertheless remains reduced under illumination after the IQM reaches a stable state. The following subsection examines the role of optical attenuation in this reduction.

\subsection{Power degradation analysis}

After ABC feedback restores stable modulation, it compensates for phase-related bias drift but does not remove the field attenuation represented by $\eta_I$ and $\eta_Q$. We treat the modulation responses $I_I$ and $I_Q$ as approximately independent of $P_b$ after relocking. At the quadrature setting $\phi=\pi/2$, the interference term in~\cref{eq:p_out} vanishes. With fixed input optical power, the assumed decrease in $\eta_I$ and $\eta_Q$ lowers the average output optical power as $P_b$ increases,
\begin{equation}
\frac{d\langle P_{\mathrm{out}}\rangle}{dP_b}<0,
\end{equation}
where the angle brackets denote averaging over the modulated signals.

Because the electrical drive and VOA setting were fixed during the experiment, reduced field transmission lowers the amplitude of the modulated optical signal and thus the modulation variance $V_A$. This mechanism explains why reduced modulation variance can coexist with stable operation and near-baseline excess-noise estimates under $4$ and $5~\mu\mathrm{W}$ illumination. After the illumination is removed, relaxation of the photoinduced perturbation allows the modulation variance to return toward its initial value.

\subsection{Possible degradation of ABC tracking performance under strong blue-light illumination}

The preceding analysis considers a photoinduced bias drift that remains within the effective compensation range of the ABC feedback loop. In this regime, ABC compensates for the bias variation and restores stable modulation. The field transmission can nevertheless remain reduced, leading to the lower modulation variance observed under stable illumination. At higher blue-light power, a larger or more rapidly varying optical-phase perturbation may approach the tracking limits of the ABC. The feedback loop may then fail to follow the changing operating point even as it continues to adjust the bias voltages. The resulting time-dependent bias variation can make the transmitted optical states unstable and may appear as additional excess noise in a CV-QKD system. This interpretation is consistent with the continued modulation-variance fluctuations observed at $6$ and $7~\mu\mathrm{W}$, although the ABC bias trajectories were not directly measured.


\begin{acknowledgments}
We thank Jialei Su and Zihao Chen for helpful discussions.
This work was supported by the National Natural Science Foundation of China (62371459, 62671369); Quantum Science and Technology-National Science and Technology Major Project (2021ZD0300703, 2021ZD0300704); Shanghai Municipal Science and Technology Major Project (2019SHZDZX01); and Hunan Provincial Natural Science Foundation (2026JJ20059).
\end{acknowledgments}

\section*{Disclosures}
The authors declare no conflicts of interest.

\section*{Data availability}
All data supporting the findings of this study are reported in the main text and appendixes. Source data and code are available from the corresponding authors upon reasonable request.


\appendix
\section{Channel-parameter estimation}
\label{app:channel_parameter_estimation}

The quantum-channel transmittance and excess noise were estimated from correlated Gaussian variables prepared by the transmitter and measured by the receiver. The two quadratures were processed independently. In the following, $x_A$ and $x_B$ denote the transmitted and received variables for either the $X$ or $P$ quadrature. The transmitted variable was calibrated at the input reference plane of the quantum channel, with
\begin{equation}
\operatorname{Var}(x_A)=V_A,
\label{eq:alice_modulation_variance}
\end{equation}
where $V_A$ is the modulation variance in SNU. The received data were normalized using the independently measured shot-noise variance,
\begin{equation}
x_B
=
\frac{x_{B,\mathrm{raw}}}{\sqrt{N_0}},
\label{eq:bob_snu_normalization}
\end{equation}
where $x_{B,\mathrm{raw}}$ denotes the measured quadrature data before SNU normalization.

After DSP, the transmitted and received variables were described by a Gaussian linear model~\cite{PhysRevA.81.062343,PhysRevA.86.032309,PhysRevA.100.012313},
\begin{equation}
x_B
=
t x_A+z,
\label{eq:linear_channel_model}
\end{equation}
where $t$ is the effective amplitude transmission coefficient and $z$ is a centered Gaussian noise variable independent of $x_A$. Under the trusted-detector model adopted here,
\begin{equation}
t=\sqrt{\eta T},
\label{eq:t_eta_T}
\end{equation}
where $T$ is the power transmittance between the quantum-channel reference planes and $\eta$ is the receiver detection efficiency. With the received quadrature normalized to shot noise, the residual-noise variance is
\begin{equation}
\sigma_z^2
=
1+v_{\mathrm{el}}+\eta T\xi,
\label{eq:residual_variance}
\end{equation}
where the unit term is the vacuum-noise contribution, $v_{\mathrm{el}}$ is the electronic noise in SNU, and $\xi$ is the total excess noise referred to the channel input. Using~\cref{eq:t_eta_T}, the same relation becomes
\begin{equation}
\sigma_z^2
=
1+v_{\mathrm{el}}+t^2\xi.
\label{eq:residual_variance_t}
\end{equation}

For a block of $m$ paired quadrature values, the sample means and centered variables are defined by
\begin{equation}
\overline{x}_A
=
\frac{1}{m}\sum_{i=1}^{m}x_{A,i},
\qquad
\overline{x}_B
=
\frac{1}{m}\sum_{i=1}^{m}x_{B,i},
\label{eq:sample_means}
\end{equation}
\begin{equation}
\delta x_{A,i}
=
x_{A,i}-\overline{x}_A,
\qquad
\delta x_{B,i}
=
x_{B,i}-\overline{x}_B.
\label{eq:centered_variables}
\end{equation}
The amplitude transmission coefficient is estimated from the covariance of the paired data,
\begin{equation}
\widehat{t}
=
\frac{
\displaystyle\sum_{i=1}^{m}
\delta x_{A,i}\delta x_{B,i}
}{
\displaystyle\sum_{i=1}^{m}
(\delta x_{A,i})^2
}.
\label{eq:t_estimator}
\end{equation}
Equivalently, writing the sample averages with angle brackets gives
\begin{equation}
\widehat{t}
=
\frac{
\left\langle\delta x_A\delta x_B\right\rangle
}{
\left\langle(\delta x_A)^2\right\rangle
}.
\label{eq:t_estimator_covariance}
\end{equation}
The residual-noise variance is then calculated as
\begin{equation}
\widehat{\sigma}_z^2
=
\frac{1}{m}
\sum_{i=1}^{m}
\left(
\delta x_{B,i}
-
\widehat{t}\,\delta x_{A,i}
\right)^2.
\label{eq:residual_estimator}
\end{equation}

The estimated channel transmittance and input-referred excess noise follow from~\cref{eq:t_eta_T,eq:residual_variance_t},
\begin{equation}
\widehat{T}
=
\frac{\widehat{t}^{\,2}}{\eta},
\label{eq:T_estimator}
\end{equation}
\begin{equation}
\widehat{\xi}
=
\frac{
\widehat{\sigma}_z^2
-
1
-
v_{\mathrm{el}}
}{
\widehat{t}^{\,2}
}.
\label{eq:xi_estimator}
\end{equation}
The experimentally calibrated electronic noise is
\begin{equation}
v_{\mathrm{el}}=0.069~\mathrm{SNU}.
\end{equation}
For this value, the excess-noise estimator can also be written as
\begin{equation}
\widehat{\xi}
=
\frac{
\widehat{\sigma}_z^2-1.069
}{
\widehat{t}^{\,2}
}.
\label{eq:xi_estimator_experimental}
\end{equation}
This procedure was applied independently to the $X$ and $P$ quadratures, and the resulting estimates were compared for consistency.

The parameter-estimation procedure above assumes that Alice's modulation calibration remains valid for the data being analyzed. In the attack scenario considered in the main text, Alice calibrates the IQM under blue-light illumination but continues to use that calibration after the device recovers. Thus, the old calibration is applied to newly generated symbols whose actual optical modulation variance has changed. Let $x_{A,0}$ and $V_{A,0}$ denote the quadrature variable and variance assigned using the illuminated-state calibration. These correspond to $x_A$ and $V_A$ in the main-text attack analysis. We model the change in the actual encoded variable as a linear rescaling. The physical channel transmittance, channel excess noise, detection efficiency, shot noise, and electronic noise are assumed to remain unchanged.

If the actual modulation variance changes by a factor $k$, then
\begin{equation}
k
=
\frac{V_A'}{V_{A,0}},
\qquad
x_A'
=
\sqrt{k}\,x_{A,0},
\label{eq:modulation_scaling_factor}
\end{equation}
where $V_A'$ and $x_A'$ describe the states encoded after the change. If Alice and Bob continue to use $x_{A,0}$ when estimating the channel from these new data, the corresponding theoretical moments are
\begin{equation}
\left\langle x_{A,0}^2\right\rangle
=
V_{A,0},
\end{equation}
\begin{equation}
\left\langle x_{A,0}x_B'\right\rangle
=
\sqrt{k\eta T}\,V_{A,0},
\label{eq:scaled_covariance}
\end{equation}
\begin{equation}
\left\langle(x_B')^2\right\rangle
=
k\eta T V_{A,0}
+
\eta T\xi
+
1
+
v_{\mathrm{el}}.
\label{eq:scaled_bob_variance}
\end{equation}

Using the original calibration, Alice and Bob interpret these same moments in terms of apparent parameters $T_{\mathrm{est}}$ and $\xi_{\mathrm{est}}$. The covariance gives
\begin{equation}
\sqrt{\eta T_{\mathrm{est}}}\,V_{A,0}
=
\sqrt{k\eta T}\,V_{A,0},
\label{eq:scaled_transmittance_relation}
\end{equation}
while the received variance gives
\begin{equation}
\eta T_{\mathrm{est}}V_{A,0}
+
\eta T_{\mathrm{est}}\xi_{\mathrm{est}}
+
1
+
v_{\mathrm{el}}
=
k\eta T V_{A,0}
+
\eta T\xi
+
1
+
v_{\mathrm{el}}.
\label{eq:scaled_noise_relation}
\end{equation}
Solving these relations yields
\begin{equation}
T_{\mathrm{est}}=kT,
\qquad
\xi_{\mathrm{est}}=\frac{\xi}{k},
\label{eq:biased_channel_parameters}
\end{equation}
consistent with Ref.~\cite{PhysRevA.100.012313}. Within this scaling model, an unrecognized increase in modulation variance makes the channel appear more transmissive and less noisy than the physical channel.

\section{405-nm optical characterization and estimation of the fiber laser-induced damage threshold}
\label{App:405_setup_LIDT}

In the experiments described in the main text, the 405-nm light was coupled into the optical path between the IQM and the VOA. This arrangement provided microwatt-level blue-light power at the IQM using the available laser. In the external injection configuration, light entering from the quantum channel would first pass through the VOA. We therefore measured the VOA attenuation at \SI{405}{\nano\meter} to determine the power required at its input, then characterized the circulator and beam splitter used in the two injection configurations.

\begin{figure}[htbp]
\centering
\includegraphics[width=0.5\linewidth]{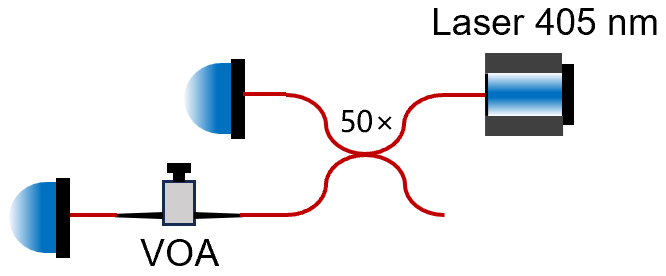}
\caption{
Arrangement used to measure the attenuation of the VOA at \SI{405}{\nano\meter}. The blue-light powers before and after the VOA were measured with a power meter while the VOA remained at the setting used in the experiment.
}
\label{Fig:appendix_setup}
\end{figure}

During Step 1 of the procedure in~\cref{experiment_procedure}, the VOA was adjusted until the modulation variance reached approximately $V_A=6.2~\mathrm{SNU}$. Its attenuation setting was then kept fixed. The measurement arrangement in~\cref{Fig:appendix_setup} compares the blue-light powers before and after the VOA. As shown in~\cref{tab:voa_isolation}, the measured attenuation at \SI{405}{\nano\meter} is approximately $57$--$58~\mathrm{dB}$. For a target power of $5~\mu\mathrm{W}$ at the IQM, this attenuation alone corresponds to approximately $2.5$--$3.2~\mathrm{W}$ at the VOA input. Losses elsewhere along an external injection path would increase the power required at Eve's launch point.

\begin{table*}[htbp]
\caption{Measured attenuation of the VOA at \SI{405}{\nano\meter}.}
\label{tab:voa_isolation}
\centering
\begin{tabular}{ccc}
\toprule
Blue-light Input (dBm)
&
Blue-light Output (dBm)
&
Attenuation (dB)
\\
\midrule
11.8 & -45.0 & 56.8 \\
13.3 & -44.2 & 57.5 \\
14.5 & -43.5 & 58.0 \\
15.4 & -42.3 & 57.7 \\
\bottomrule
\end{tabular}
\end{table*}

To compare the VOA input-power requirement with the power-handling capability of optical fiber, we consider a nominal estimate of the laser-induced damage threshold (LIDT) under continuous-wave illumination. When LIDT is expressed as a threshold irradiance, the corresponding optical-power threshold can be estimated as
\begin{equation}
P_{\max}
=
\mathrm{LIDT}\cdot A_{\mathrm{eff}},
\end{equation}
where $A_{\mathrm{eff}}$ is the effective area of the guided optical mode. Ref.~\cite{PhysRevX.5.031030} uses a nominal continuous-wave value of
\begin{equation}
P_{\max}(1550~\mathrm{nm})=12.8~\mathrm{W}
\end{equation}
in an illustrative fiber-damage analysis. That work also discusses an approximate square-root dependence on wavelength~\cite{PhysRevLett.91.127402},
\begin{equation}
\frac{\mathrm{LIDT}(\lambda_1)}
{\mathrm{LIDT}(\lambda_2)}
\approx
\sqrt{\frac{\lambda_1}{\lambda_2}}.
\end{equation}
Formally applying this relation from 1550 to \SI{405}{\nano\meter}, while holding the effective area fixed, gives
\begin{equation}
P_{\mathrm{nom}}(405~\mathrm{nm})
=
12.8~\mathrm{W}
\sqrt{\frac{405}{1550}}
\approx
6.54~\mathrm{W}.
\end{equation}

The value \SI{6.54}{W} is a nominal extrapolation, not a measured continuous-wave damage threshold for the fiber and components in the proposed 405-nm injection path. Its comparison with the VOA input requirement indicates the scale of the optical powers involved, but does not establish damage-free transmission through the full path. Injection through the quantum channel may be feasible with optical components selected for sufficiently low loss and adequate power handling at \SI{405}{\nano\meter}.

The commercial 1550-nm circulator shown in the external attack configuration was also characterized at \SI{405}{\nano\meter}. Its output power remained below the measurement limit of $-60~\mathrm{dBm}$ for each tested input power, giving the attenuation lower bounds in~\cref{tab:BsCIR_isolation}. The 50:50 beam splitter used for the experimental illumination path had substantially lower measured attenuation of approximately $3.6$--$4.3~\mathrm{dB}$. This difference explains the use of the beam splitter in the experiment. The measured circulator losses characterize that particular 1550-nm component and do not determine the loss of every possible external injection path at \SI{405}{\nano\meter}.

\begin{table*}[htbp]
\caption{Measured attenuation of the optical circulator and 90:10 beam splitter at \SI{405}{\nano\meter}. Circulator attenuation values are lower bounds because its output power was below the measurement limit.}
\label{tab:BsCIR_isolation}
\centering
\begin{tabular}{ccccc}
\toprule
\multirow{2}{*}{\shortstack{Blue-light\\Input (dBm)}}
&
\multicolumn{2}{c}{Optical Circulator}
&
\multicolumn{2}{c}{Beam Splitter}
\\
\cmidrule(lr){2-3}
\cmidrule(lr){4-5}
&
\shortstack{Output\\Power (dBm)}
&
\shortstack{Attenuation\\(dB)}
&
\shortstack{Output\\Power (dBm)}
&
\shortstack{Attenuation\\(dB)}
\\
\midrule
11.8 & $<-60$ & $>71.8$ & 8.2 & 3.6 \\
13.3 & $<-60$ & $>73.3$ & 9.4 & 3.9 \\
14.5 & $<-60$ & $>74.5$ & 10.4 & 4.1 \\
15.4 & $<-60$ & $>75.4$ & 11.1 & 4.3 \\
\bottomrule
\end{tabular}
\end{table*}

\section{Digital signal processing procedure}
\label{App:DSP}

DSP is used to prepare the transmitted modulation waveform and recover the received coherent-detection data. Because the transmitter and receiver use independent lasers, their relative phase changes over time. A frequency-division-multiplexed pilot tone provides the phase reference needed to process the quantum signal. The pilot and quantum signal occupy separate frequency bands and are separated digitally at the receiver. The processing procedure is summarized below.

\textbf{(1) Generation of Gaussian-modulated quantum symbols.}
Independent Gaussian random sequences are generated for the in-phase and quadrature components. Each sequence is normalized to zero mean and unit variance before being scaled to the desired modulation variance. The two sequences are combined into complex-valued GMCS symbols. At a symbol rate of \SI{800}{MBaud}, a waveform lasting \SI{100}{\micro\second} contains 80000 Gaussian-modulated symbols.

\textbf{(2) Pulse shaping using an RRC filter.}
The symbol sequence is interpolated and pulse-shaped with a root-raised-cosine (RRC) filter. This limits its spectral occupancy and suppresses intersymbol interference while producing a waveform suitable for electro-optic modulation. The RRC filter used here has a roll-off factor of $0.1$ and a span of 12 symbol periods.

\textbf{(3) Digital intermediate-frequency up-conversion.}
After pulse shaping, the quantum signal is shifted from baseband to an intermediate frequency of \SI{500}{MHz}. This placement leaves spectral space for the pilot tone and lies within the bandwidth of the electrical modulation and acquisition chain.

\textbf{(4) Frequency-division multiplexing of the pilot signal.}
A continuous-wave sinusoidal pilot tone is placed at \SI{-1}{GHz}. The pilot and quantum signal are combined digitally before conversion to the I/Q electrical waveforms delivered by the AWG. Their amplitude ratio is fixed at $20$.

\textbf{(5) Acquisition of the received I/Q signals.}
After optical transmission and coherent detection, the electrical outputs of the two balanced detectors are sampled by the oscilloscope. The sampled in-phase and quadrature channels are combined into a complex-valued received waveform. This waveform contains both the quantum signal and the pilot component used for phase recovery.

\textbf{(6) Pilot extraction and phase estimation.}
The received waveform is down-converted at the pilot frequency to move the pilot component to baseband. A digital low-pass filter extracts this component, and its instantaneous phase is calculated over time. Pilot extraction uses a filter with a cutoff bandwidth of \SI{2}{MHz} and an order of 801. The pilot tracks the relative laser-frequency offset and phase drift experienced by the received signal. Its extracted phase is unwrapped and further smoothed with a moving-average filter.

\textbf{(7) Quantum-signal phase compensation.}
The quantum signal is down-converted from its \SI{500}{MHz} intermediate frequency to baseband. The phase estimated from the pilot is then applied as a complex rotation to compensate for the relative phase variation between the transmitter and receiver lasers. A subsequent digital low-pass filter suppresses out-of-band noise. The quantum-signal filter has an order of 2401, with its cutoff selected according to the \SI{800}{MBaud} symbol rate and RRC roll-off factor.

\textbf{(8) Matched filtering and clock recovery.}
The phase-compensated signal is passed through an RRC matched filter with the transmitter's filter parameters. A publicly disclosed portion of the transmitted symbol sequence serves as a training sequence for clock recovery. Sampling offsets within one symbol period are searched, and the offset producing the greatest correlation with the training sequence is selected. The recovered symbols are then available for channel-parameter estimation and excess-noise evaluation. The resulting parameters are used in the asymptotic key-rate calculations described in the following section.

\section{Asymptotic secret key rate for heterodyne detection under collective attacks}
\label{App:keyrate_sim}

We use the asymptotic secret-key-rate formulation for reverse-reconciled CV-QKD with heterodyne detection under collective attacks. The rate per transmitted pulse is
\begin{equation}
R
=
\beta I_{AB}^{\mathrm{het}}
-
\chi_{BE}^{\mathrm{het}},
\end{equation}
where $\beta$ is the reconciliation efficiency, $I_{AB}^{\mathrm{het}}$ is the mutual information between Alice and Bob, and $\chi_{BE}^{\mathrm{het}}$ is the Holevo information between Eve and Bob. A positive value of $R$ corresponds to a positive asymptotic key rate.

For heterodyne detection, Alice and Bob's mutual information is
\begin{equation}
I_{AB}^{\mathrm{het}}
=
\log_2
\left(
\frac{
V+\chi_{\mathrm{tot}}
}{
1+\chi_{\mathrm{tot}}
}
\right).
\end{equation}
The total input-referred noise and its channel and detection contributions are
\begin{equation}
\begin{aligned}[c]
\chi_{\mathrm{tot}}
&=
\chi_{\mathrm{line}}
+
\frac{\chi_{\mathrm{het}}}{T},\\
\chi_{\mathrm{line}}
&=
\frac{1}{T}-1+\xi,\\
\chi_{\mathrm{het}}
&=
\frac{1+(1-\eta)+2v_{\mathrm{el}}}{\eta}.
\end{aligned}
\end{equation}
Here,
\begin{equation}
V=V_A+1,
\end{equation}
where $V_A$ is Alice's modulation variance in SNU and the additional unit represents vacuum noise. The parameters $T$ and $\xi$ denote the channel transmittance and input-referred excess noise. The trusted heterodyne receiver is described by its detection efficiency $\eta$ and electronic noise $v_{\mathrm{el}}$.

The Holevo information is evaluated from the symplectic eigenvalues,
\begin{equation}
\chi_{BE}^{\mathrm{het}}
=
\sum_{i=1}^{2}
G\left(
\frac{\lambda_i-1}{2}
\right)
-
\sum_{i=3}^{5}
G\left(
\frac{\lambda_i-1}{2}
\right),
\end{equation}
with
\begin{equation}
G(x)
=
(x+1)\log_2(x+1)-x\log_2x.
\end{equation}
The first two symplectic eigenvalues are
\begin{equation}
\lambda_{1,2}^{2}
=
\frac{1}{2}
\left(
A\pm\sqrt{A^2-4B}
\right),
\end{equation}
while the remaining eigenvalues satisfy
\begin{equation}
\begin{aligned}[c]
\lambda_{3,4}^{2}
&=
\frac{1}{2}
\left(
C\pm\sqrt{C^2-4D}
\right),\\
\lambda_5&=1.
\end{aligned}
\end{equation}
The quantities $A$, $B$, $C$, and $D$ are defined by
\begin{equation}
\begin{aligned}[c]
A
&=
V^2(1-2T)+2T
+
T^2(V+\chi_{\mathrm{line}})^2,\\
B
&=
T^2(V\chi_{\mathrm{line}}+1)^2,
\end{aligned}
\end{equation}
\begin{equation}
\begin{aligned}[c]
C
&=
\frac{1}{[T(V+\chi_{\mathrm{tot}})]^2}
\Bigl[
A\chi_{\mathrm{het}}^2+B+1\\
&\quad
+
2\chi_{\mathrm{het}}
\bigl(
V\sqrt{B}+T(V+\chi_{\mathrm{line}})
\bigr)
+
2T(V^2-1)
\Bigr],
\end{aligned}
\end{equation}
\begin{equation}
D
=
\left[
\frac{
V+\sqrt{B}\chi_{\mathrm{het}}
}{
T(V+\chi_{\mathrm{tot}})
}
\right]^2.
\end{equation}

These expressions are used with the measured modulation and noise parameters to calculate the asymptotic rates discussed in the main text. The comparison between rates calculated from the illuminated-state calibration and from the recovered modulation variance follows the attack scenario defined in Appendix~\ref{app:channel_parameter_estimation}.


\begin{thebibliography}{44}
\bibitem [{not()}]{note:effective_loss}%
  \BibitemOpen
  \href@noop {} {}\bibinfo {note} {The mean transmittance $T=0.065$ inferred from the Alice--Bob data corresponds to an effective attenuation of approximately $11.9~\mathrm{dB}$. This exceeds the directly measured optical-channel attenuation of approximately $10.2~\mathrm{dB}$. The difference may reflect imperfect polarization or temporal-mode matching and residual phase-compensation errors.}\BibitemShut {Stop}%
\end{thebibliography}
\end{document}